\documentclass[conference]{IEEEtran}
\IEEEoverridecommandlockouts
\usepackage{cite}
\usepackage{amsmath,amssymb,amsfonts}
\usepackage{algorithm}
\usepackage{graphicx}
\usepackage{textcomp}
\usepackage{xcolor}
\usepackage{todonotes}

\usepackage{multirow}
\usepackage{comment}
\usepackage{verbatim} 
\newcommand{\xmark}{\ding{55}}
\usepackage{tikz}
\usepackage{pgfplots}
\usepackage{pgfplotstable}
\usepackage{graphicx}
\usepackage{multirow}
\usepackage{pifont}
\usepackage[export]{adjustbox}
\usepackage{pgf-pie}
\usepackage{tkz-graph}
\usepackage{graphicx}
\usepackage{subcaption}
\usepackage[skip=2pt]{caption}
\usepackage{hyperref}
\usepackage{microtype}
\usepackage{algorithm}
\usepackage{algpseudocode}
\usepackage{xcolor} 

\newcommand*\circled[1]{%
  \tikz[baseline=(C.base)]{
    \node[draw, circle, inner sep=1pt, minimum size=0.8em] (C) {#1};
  }%
}

\newcommand{\cmark}{\ding{51}}

\definecolor{SlateBlue}{RGB}{106, 90, 205} 
\definecolor{CadetBlue}{RGB}{95, 158, 160} 
\definecolor{BurntOrange}{RGB}{204, 85, 0} 
\definecolor{Goldenrod}{RGB}{218, 165, 32} 

\hypersetup{
    colorlinks,
    linkcolor={red!40!black},
    citecolor={blue!40!black},
    urlcolor={black}
}
\usepackage{booktabs}
\usepackage{titlesec}

\usetikzlibrary{pgfplots.groupplots}
\usetikzlibrary{shadows,patterns,shapes,arrows,decorations.pathmorphing,backgrounds,positioning,fit,plotmarks,calc,spy,matrix}
\pgfplotsset{compat=1.11,
    /pgfplots/ybar legend/.style={
    /pgfplots/legend image code/.code={%
       \draw[##1,/tikz/.cd,yshift=-0.25em]
        (0cm,0cm) rectangle (3pt,0.8em);},
   },
}
\usetikzlibrary{decorations.text}
\definecolor{cadetblue}{rgb}{0.37, 0.62, 0.63}
\definecolor{airforceblue}{rgb}{0.36, 0.54, 0.66}
\definecolor{caribbeangreen}{rgb}{0.0, 0.8, 0.6}
\definecolor{carolinablue}{rgb}{0.6, 0.73, 0.89}
\definecolor{darkgoldenrod}{rgb}{0.72, 0.53, 0.04}
\definecolor{debianred}{rgb}{0.84, 0.04, 0.33}
\definecolor{fuzzywuzzy}{rgb}{0.8, 0.4, 0.4}
\definecolor{grullo}{rgb}{0.66, 0.6, 0.53}
\definecolor{ceil}{rgb}{0.57, 0.63, 0.81}
\definecolor{candypink}{rgb}{0.89, 0.44, 0.48}
\definecolor{calpolypomonagreen}{rgb}{0.12, 0.3, 0.17}
\definecolor{burntsienna}{rgb}{0.91, 0.45, 0.32}
\definecolor{atomictangerine}{rgb}{1.0, 0.6, 0.4}
\definecolor{goldenrod}{rgb}{0.85, 0.65, 0.13}
\definecolor{gamboge}{rgb}{0.89, 0.61, 0.06}
\definecolor{amber}{rgb}{1.0, 0.75, 0.0}
\definecolor{battleshipgrey}{rgb}{0.52, 0.52, 0.51}
\definecolor{darkcerulean}{rgb}{0.03, 0.27, 0.49}
\definecolor{fuzzywuzzy}{rgb}{0.8, 0.4, 0.4}
\definecolor{mediumseagreen}{rgb}{0.24, 0.7, 0.44}
\definecolor{antiquebrass}{rgb}{0.8, 0.58, 0.46}
\definecolor{apricot}{rgb}{0.98, 0.81, 0.69}
\definecolor{asparagus}{rgb}{0.53, 0.66, 0.42}
\definecolor{bananamania}{rgb}{0.98, 0.91, 0.71}
\definecolor{cadmiumgreen}{rgb}{0.0, 0.42, 0.24}
\definecolor{chocolate}{rgb}{0.48, 0.25, 0.0}
\definecolor{cinereous}{rgb}{0.6, 0.51, 0.48}
\definecolor{aliceblue}{rgb}{0.94, 0.97, 1.0}
\definecolor{beaublue}{rgb}{0.74, 0.83, 0.9}
\definecolor{blizzardblue}{rgb}{0.67, 0.9, 0.93}
\definecolor{bittersweet}{rgb}{1.0, 0.44, 0.37}
\definecolor{camouflagegreen}{rgb}{0.47, 0.53, 0.42}
\definecolor{darkolivegreen}{rgb}{0.33, 0.42, 0.18}
\definecolor{darkpastelblue}{rgb}{0.47, 0.62, 0.8}
\definecolor{desertsand}{rgb}{0.93, 0.79, 0.69}
\definecolor{deeppeach}{rgb}{1.0, 0.8, 0.64}
\definecolor{indianred}{rgb}{0.8, 0.36, 0.36}
\definecolor{oldmauve}{rgb}{0.4, 0.19, 0.28}
\definecolor{lightblue}{rgb}{0.68, 0.85, 0.9}
\definecolor{lightcyan}{rgb}{0.88, 1.0, 1.0}
\definecolor{viridian}{rgb}{0.25, 0.51, 0.43}
\definecolor{slategray}{rgb}{0.44, 0.5, 0.56}
\definecolor{manatee}{rgb}{0.59, 0.6, 0.67}
\definecolor{darkbrown}{rgb}{0.4, 0.26, 0.13}
\definecolor{almond}{rgb}{0.94, 0.87, 0.8}
\def\BibTeX{{\rm B\kern-.05em{\sc i\kern-.025em b}\kern-.08em
    T\kern-.1667em\lower.7ex\hbox{E}\kern-.125emX}}
\usepackage{colortbl}
\usepackage{tabulary}
\pgfkeys{%
/piechartthreed/.cd,
scale/.code                =  {\def\piechartthreedscale{#1}},
mix color/.code            =  {\def\piechartthreedmixcolor{#1}},
background color/.code     =  {\def\piechartthreedbackcolor{#1}},
name/.code                 =  {\def\piechartthreedname{#1}}}
\newcommand\piechartthreed[2][]{%
   \pgfkeys{/piechartthreed/.cd,
     scale            = 1,
     mix color        = gray,
     background color = white,
     name             = pc} 
  \pgfqkeys{/piechartthreed}{#1}
  \begin{scope}[scale=\piechartthreedscale] 
  \begin{scope}[xscale=5,yscale=3] 
     \path[preaction={fill=black,opacity=.8,
         path fading=circle with fuzzy edge 20 percent,
         transform canvas={yshift=-15mm*\piechartthreedscale}}] (0,0) circle (1cm);
     \pgfmathsetmacro\totan{0} 
     \global\let\totan\totan 
     \pgfmathsetmacro\bottoman{180} \global\let\bottoman\bottoman 
     \pgfmathsetmacro\toptoman{0}   \global\let\toptoman\toptoman 
     \begin{scope}[draw=black,thin]
     \foreach \an/\col [count=\xi] in {#2}{%
     \def\space{ } 
        \coordinate (\piechartthreedname\space\xi) at (\totan+\an/2:0.75cm); 
        \ifdim 180pt>\totan pt 
         \ifdim 0pt=\toptoman pt
            \pgfmathsetmacro\toptoman{180} 
            \global\let\toptoman\toptoman         
            \else
          \fi
        \fi   
        \fill[\col!80!gray,draw=black] (0,0)--(\totan:1cm)  arc(\totan:\totan+\an:1cm)
                                     --cycle;     
       \pgfmathsetmacro\finan{\totan+\an}
       \ifdim 180pt<\finan pt 
         \ifdim 180pt=\bottoman pt
            \shadedraw[left color=\col!20!\piechartthreedmixcolor,
                       right color=\col!5!\piechartthreedmixcolor,
                       draw=black,very thin] (180:1cm) -- ++(0,-3mm) arc (180:\totan+\an:1cm) 
                                                       -- ++(0,3mm)  arc (\totan+\an:180:1cm);
            \pgfmathsetmacro\bottoman{0}
            \global\let\bottoman\bottoman
            \else
            \shadedraw[left color=\col!20!\piechartthreedmixcolor,
                       right color=\col!5!\piechartthreedmixcolor,
                       draw=black,very thin](\totan:1cm)-- ++(0,-3mm) arc(\totan:\totan+\an:1cm)
                                                        -- ++(0,3mm)  arc(\totan+\an:\totan:1cm); 
          \fi
        \fi
        \pgfmathsetmacro\totan{\totan+\an}  \global\let\totan\totan 
       } 
    \end{scope}
   \end{scope}  
\end{scope}
}
\def\BibTeX{{\rm B\kern-.05em{\sc i\kern-.025em b}\kern-.08em
    T\kern-.1667em\lower.7ex\hbox{E}\kern-.125emX}}

\begin{document}

\makeatletter
\newcommand\bstctlcite@aux[1]{%
  \if@filesw
    \immediate\write\@auxout{\string\bstctlcite{#1}}%
  \fi}
\makeatother
\bstctlcite{CTLmaxauthors}

\title{\vspace{-1em}{\footnotesize\textit{Invited Paper}}\\[0.5em]Building an Open CGRA Ecosystem\\ for Agile Innovation}

    \makeatletter
    \newcommand{\linebreakand}{%
      \end{@IEEEauthorhalign}
      \hfill\mbox{}\par
      \mbox{}\hfill\begin{@IEEEauthorhalign}
    }
    \makeatother

\author{\IEEEauthorblockN{Rohan Juneja}
\IEEEauthorblockA{\textit{National University of Singapore} \\
rohan@comp.nus.edu.sg}
\and
\IEEEauthorblockN{Pranav Dangi}
\IEEEauthorblockA{\textit{National University of Singapore} \\
dangi@comp.nus.edu.sg}
\and
\IEEEauthorblockN{Thilini Kaushalya Bandara}
\IEEEauthorblockA{\textit{Renesas Electronics} \\
thilini@comp.nus.edu.sg}
\linebreakand
\IEEEauthorblockN{Zhaoying Li}
\IEEEauthorblockA{\textit{National University of Singapore} \\
zhaoying@comp.nus.edu.sg}
\and
\IEEEauthorblockN{Dhananjaya Wijerathne}
\IEEEauthorblockA{\textit{Advanced Micro Devices} \\
dmd.wijerathne@amd.com}
\linebreakand
\IEEEauthorblockN{Li-Shiuan Peh}
\IEEEauthorblockA{\textit{National University of Singapore} \\
peh@comp.nus.edu.sg}
\and
\IEEEauthorblockN{Tulika Mitra}
\IEEEauthorblockA{\textit{National University of Singapore} \\
tulika@comp.nus.edu.sg}}


\maketitle

\begin{abstract}
Modern computing workloads, particularly in AI and edge applications, demand hardware-software co-design to meet aggressive performance and energy targets. Such co-design benefits from open and agile platforms that replace closed, vertically integrated development with modular, community-driven ecosystems. Coarse-Grained Reconfigurable Architectures (CGRAs), with their unique balance of flexibility and efficiency, are particularly well-suited for this paradigm. When built on open-source hardware generators and software toolchains, CGRAs provide a compelling foundation for architectural exploration, cross-layer optimization, and real-world deployment.

In this paper, we will present an open CGRA ecosystem that we have developed to support agile innovation across the stack. Our contributions include HyCUBE, a CGRA with a reconfigurable single-cycle multi-hop interconnect for efficient data movement; PACE, which embeds a power-efficient HyCUBE within a RISC-V SoC targeting edge computing; and Morpher, a fully open-source, architecture-adaptive CGRA design framework that supports design space exploration, compilation, simulation, and validation. By embracing openness at every layer, we aim to lower barriers to innovation, enable reproducible research, and demonstrate how CGRAs can anchor the next wave of agile hardware development. We will conclude with a call for a unified abstraction layer for CGRAs and spatial accelerators, one that decouples hardware specialization from software development. Such a representation would unlock architectural portability, compiler innovation, and a scalable, open foundation for spatial computing.
\end{abstract}


\section{Introduction}



Modern workloads spanning large-scale deep learning inference (from convolutional neural networks in vision to transformer-based language models), real-time sensor processing in autonomous systems, and ultra-low-power analytics at the edge demand a unique combination of high throughput, tight latency bounds, and extreme energy efficiency (performance-per-watt) that general-purpose processors cannot deliver. Domain-specific accelerators (DSAs) have emerged as an efficient solution for such workloads and are now pervasive in modern SoCs~\cite{sigarch_mobile_socs}; for instance, Shao et al.~\cite{shao_dsa} report that Apple’s SoCs have grown from 10 DSAs in the A4 to over 40 in the A12. However, such specialization leads to dark silicon, as many DSAs remain underutilized across diverse workloads. These applications also exhibit varied computational patterns from dense linear algebra to irregular, data-dependent kernels each requiring distinct dataflows and memory access strategies.

\textbf{Existing ecosystems are siloed:}
\textit{The prevailing hardware-software stacks remain tightly coupled and closed, limiting the flexibility and adaptability required for modern workloads}. Commercial GPUs (e.g., NVIDIA’s CUDA‑driven ecosystem with cuDNN and TensorRT), domain‑specific ASICs (e.g., Google’s TPU with XLA/TensorFlow integration), FPGA‑based dataflow (e.g., Maxeler’s proprietary MaxCompiler and runtime), and adaptive compute platforms (e.g., AMD's Versal ACAP with Vitis AI DPU IP) all represent monolithic, vendor‑controlled ecosystems. Supporting the diversity in applications is challenging for fixed-function accelerators.


While GPUs and ASICs deliver high throughput and low latency for specific workloads, they limit architectural flexibility, offer closed compilers and rigid execution models, and restrict access to micro-architectural details. FPGA-based platforms offer fine-grained reconfiguration, but still face barriers such as long compilation times due to gate-level synthesis/bit-level reconfiguration and reliance on vendor-specific toolchains.
Across all these platforms, inflexible integration and closed‑source compilation flows hide critical micro-architectural behavior, hinder third‑party compiler innovations, and lock developers into a single vendor’s framework. Consequently, architectural exploration is stifled, cross‑platform portability is hindered, and tool fragmentation presents hurdles to reproducible, community‑driven research. This dynamic elevates vendors to first‑class citizens and forces engineers to conform to predefined application profiles. It also amplifies the "hardware lottery"~\cite{hardware_lottery} effect by limiting novel research to the narrow subset of ideas compatible with existing hardware.

\textbf{Why Open and Modular platforms?}
Open and modular ecosystems decouple architectural innovation from monolithic toolchains by offering well-defined interfaces such as an architecture description language (ADL) that allows rapid customization of instruction sets or micro-architectural components without re-engineering the entire compiler stack. These abstractions let hardware architects, compiler developers, and software engineers to work independently, enabling quick and iterative experimentation and validation of new scheduling heuristics, memory hierarchies, or micro-architectural blocks. Configurable platforms shorten this feedback loop by turning infrastructure tweaks or new workload adoption with measurable performance or power gains within hours rather than weeks.
This modularity reduces dependence on proprietary toolchains, democratizes hardware-software co-design, and fosters a vibrant, community‑driven innovation cycle.

\textbf{Contributions:} In our efforts, we develop and provide an open, modular platform, Morpher~\cite{morpher_woset}, that lowers the barrier to CGRA innovation and exploration. Morpher allows researchers to quickly describe new CGRA architectures, compile real applications to them, and validate behavior and evaluate performance. It standardizes the interface between the architecture and compiler, allowing new compositions of processing elements, interconnects, or memory layouts to be explored without rebuilding the toolchain. This enables designs to move from idea to simulation and then to silicon with lowered non-recurring engineering costs. We designed and fabricated open-source HyCUBE CGRA~\cite{hycube_tapeout} and PACE~ SoC with CGRA~\cite{pace}. We release the framework, configurations, and benchmarks to enable fair, reproducible comparisons across dataflow architectures and to enable further research and iteration over our existing work.

\section{Background}
\begin{figure}[t!]
    \centering
    \resizebox{\columnwidth}{!} {
    \includegraphics[width=\textwidth]{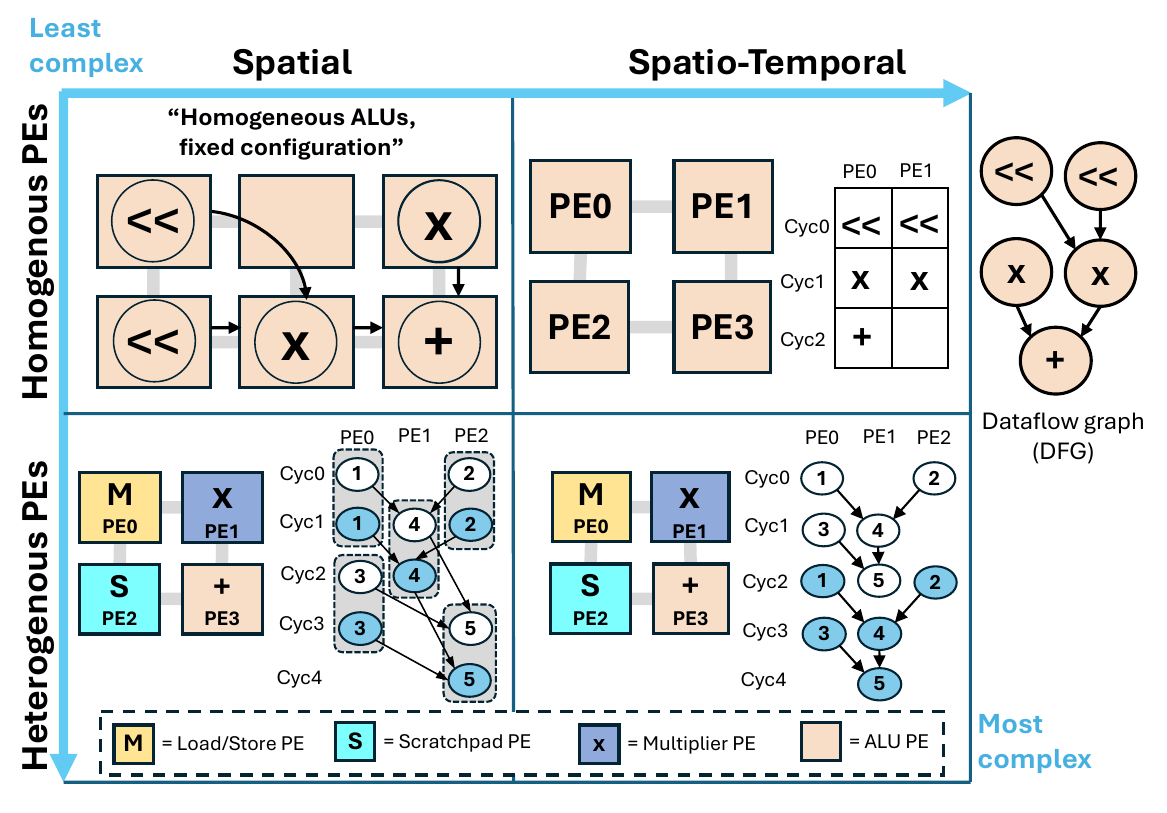}
    }
    \caption{CGRA Taxonomy.}
    \label{fig:taxonomy}
\end{figure}
CGRAs are arrays of programmable processing elements (PEs), each comprising an arithmetic logic unit, a router, and local reconfiguration memory, interconnected by a configurable fabric that maps high‑level dataflow graphs at instruction granularity. Data moves between a shared scratchpad memory and the PE array for execution, and results return to the scratchpad on completion. Unlike FPGAs, which configure lookup tables and wires at bit-level to implement arbitrary logic, or domain‑specific accelerators, which hard‑wire design for specific narrow set of kernels, CGRAs configure functional units and network paths to flexibly support a broad range of workloads with near‑ASIC efficiency. 

There has been extensive exploration of CGRA architectures from both commercial and academic groups. 
To organize the design space, we introduce a taxonomy as shown in Fig.~\ref{fig:taxonomy}. The vertical axis contrasts homogeneous arrays of identical PEs with heterogeneous fabrics of specialized units, and the horizontal axis spans spatial mappings (where each operation occupies a PE for its full execution) to spatio-temporal schedules that time multiplex operations across PEs.
In the spatial case, the entire DFG can fit onto the CGRA resources without time multiplexing. If the DFG does not fit, the loop body is split into subgraphs, completing all iterations of one subgraph before moving to the next. In the spatio-temporal case, the DFG is mapped across space and time, reconfiguring PEs and interconnects every cycle. Homogeneous arrays are simpler to place and route, while heterogeneous arrays can be more area and energy efficient for matched kernels but are harder to schedule. Overall complexity increases from top left to bottom right.

In Table~\ref{tab:cgra_architectures}, each quadrant includes representative CGRA designs. Moreover, whereas most CGRAs rely on static routing determined at compile time, architectures such as MTIA~\cite{mtia}, Nexus Machine~\cite{nexus}, Canon~\cite{canon} support dynamic routing for reconfiguring data paths at runtime, accommodating irregular communication patterns for sparse and graph workloads, enhancing adaptability.

\begin{table}[t]
\centering
\renewcommand{\arraystretch}{0.8}
\resizebox{0.9\columnwidth}{!}{%
\begin{tabular}{|>{\centering\arraybackslash}m{0.5cm}|
                >{\raggedright\arraybackslash}p{2.5cm}|
                >{\raggedright\arraybackslash}p{2.5cm}|} \hline
\textbf{} & \textbf{Spatial} & \textbf{Spatio-Temporal} \\\hline

\vspace{0.2cm}\rotatebox[origin=c]{90}{\textbf{Homogenous}} \vspace{0.2cm}&  
\vspace{-1cm} MTIA~\cite{mtia}, Tensor Core~\cite{tensor}, Flex~\cite{flex} & 
\vspace{-1cm} Renesas DRP~\cite{renesas_drp}, Samsung ULP-SRP~\cite{ulp_srp}, HyCUBE~\cite{hycube}, PACE~\cite{pace}, Nexus Machine~\cite{nexus}, Canon~\cite{canon}\\\hline

\vspace{0.2cm} \rotatebox[origin=c]{90}{\textbf{Heterogenous}} \vspace{0.2cm}&
\vspace{-1cm}Warp~\cite{warp}, FPCA~\cite{fpca}, Softbrain~\cite{softbrain}, Tartan~\cite{tartan}, Piperench~\cite{piperench}, Snafu~\cite{snafu}, Riptide~\cite{riptide} &
\vspace{-1cm}Sambanova RDU~\cite{sambanova_sn40l}, Revamp~\cite{revamp}, Plaid~\cite{plaid}, Amber~\cite{amber} \\\hline
\end{tabular}%
}
\caption{Dataflow architectures classified within the CGRA taxonomy. Accelerators often fall under the umbrella of spatial dataflow architectures albeit with minimal reconfiguration.}
\label{tab:cgra_architectures}
\end{table}

Several open-source CGRA frameworks have been proposed that support modeling, mapping, and evaluation across these features, including CGRA-ME~\cite{cgra_me}, Pillars~\cite{pillars}, OpenCGRA~\cite{opencgra}, and CCF~\cite{ccf}. CGRA-ME provides compiler and RTL for traditional spatio-temporal homogeneous CGRAs but targets simple kernels, lacks control divergence, and only partly handles recurrences; it also omits detailed memory modeling and has no open-source simulator. Pillars adds a Scala description, automatic RTL, and cycle-accurate simulation, but uses CGRA-ME as its frontend and inherit those limitations. OpenCGRA and CCF support control divergence and recurrences, yet their mappers are not architecture adaptive and often need code changes to retarget new interconnects or PE layouts. OpenCGRA's simulator rely on user-written test benches and do not automate test data generation or end-to-end checks. Morpher is open source and automated. It compiles real kernels with control divergence and recurrences, models custom interconnects and memory systems, and uses an architecture-adaptive mapper. It integrates LLVM-based DFG generation and data layout and often yields lower II and faster compile times. It also includes cycle-accurate simulation and automated checking, removing manual benches. Table~\ref{tab:cgra_frameworks} compares these frameworks across control, recurrences, adaptivity, interconnects, memory modeling, simulation and validation, and RTL readiness. To our knowledge, it is the only open-source CGRA framework validated end-to-end on fabricated silicon.
\begin{table}[t]
\centering
\resizebox{\columnwidth}{!}{
\setlength{\tabcolsep}{6pt}
\begin{tabular}{|l|l|c|c|c|c|c|}
\hline
\multicolumn{2}{|c|}{\textbf{Features}} &
\rotatebox[origin=c]{90}{\textbf{CGRA-ME}} &
\rotatebox[origin=c]{90}{\textbf{Pillars}} &
\rotatebox[origin=c]{90}{\textbf{OpenCGRA}} &
\rotatebox[origin=c]{90}{\textbf{CCF}} &
\rotatebox[origin=c]{90}{\textbf{Morpher}} \\
\hline
\multirow{2}{*}{DFG Generation}
  & Models control divergence           & \xmark & \xmark & \cmark & \cmark & \cmark \\ \cline{2-7}
  & Recurrence edges                    & \xmark & \xmark & \cmark & \cmark & \cmark \\ \hline
\multirow{3}{*}{Architecture Modeling}
  & Adapt user defined architectures    & \cmark & \cmark & \cmark & \xmark & \cmark \\ \cline{2-7}
  & Multi-hop connections               & \xmark & \xmark & \xmark & \xmark & \cmark \\ \cline{2-7}
  & Different memory organizations      & \xmark & \xmark & \cmark & \xmark & \cmark \\ \hline
\multirow{3}{*}{P\&R Mapper}
  & Architecture adaptive mapping       & \cmark & \cmark & \xmark & \xmark & \cmark \\ \cline{2-7}
  & Data layout aware mapping           & \xmark & \xmark & \xmark & \xmark & \cmark \\ \cline{2-7}
  & Recurrence aware mapping            & \xmark & \xmark & \cmark & \xmark & \cmark \\ \hline
\multirow{3}{*}{Simulation \& validation}
  & Cycle accurate simulation           & \xmark & \cmark & \cmark & \cmark & \cmark \\ \cline{2-7}
  & Test data generation                & \xmark & \xmark & \xmark & \xmark & \cmark \\ \cline{2-7}
  & Validation against test data        & \xmark & \xmark & \xmark & \xmark & \cmark \\ \hline
\end{tabular}
}
\caption{Morpher versus open-source CGRA frameworks.\protect\footnotemark}
\label{tab:cgra_frameworks}
\end{table}
\footnotetext{From Morpher~\cite{morpher_woset}.}
\section{Open CGRA Infrastrcture}
Building on the need for a unified, extensible CGRA toolchain, we have developed Morpher, an open‑source, end‑to‑end CGRA framework that automates architecture generation, mapping, simulation, and validation for both homogeneous and heterogeneous designs. Leveraging Morpher, we have taped out HyCUBE, an 4×4 CGRA featuring a reconfigurable single‑cycle, multi‑hop mesh that delivers state‑of‑the‑art energy efficiency and throughput, and PACE, a modular RISC‑V edge SoC embedding a 8x8 HyCUBE alongside a memory hierarchy and bus interfaces. These silicon prototypes demonstrate Morpher’s capability to drive rapid hardware‑software co‑design. The modular structure of Morpher has also led to the development of several novel CGRA architectures, each reusing certain components from Morpher while adding new components to match their own architectural needs.
\subsection{Morpher}
\begin{figure}[h]
    \centering
    \resizebox{\columnwidth}{!} {
    \includegraphics[width=\textwidth]{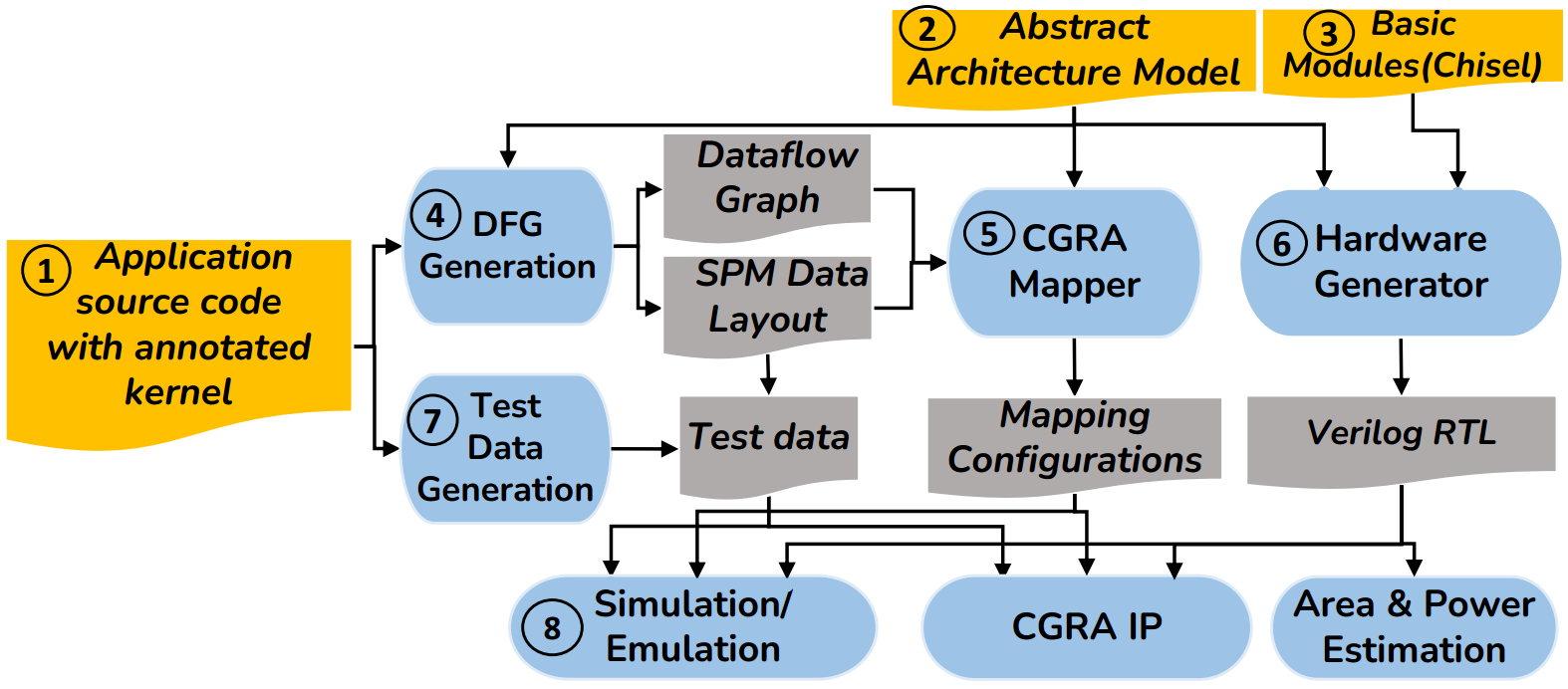}
    }
    \caption{Morpher framework\protect\footnotemark[\value{footnote}]}
    \label{fig:morpher_overview}
\end{figure}

Fig.~\ref{fig:morpher_overview} illustrates Morpher’s end‑to‑end flow, structured into three main phases.
In the \textbf{Architectural Specification \& Data‑Flow Extraction} phase, Morpher takes as input \circled{1} application source code annotated with the target kernel, \circled{2} an ADL description of the CGRA, and \circled{3} a library of Chisel hardware primitives. It then generates the Data‑Flow Graph and scratchpad memory layout (\circled{4}) and produces test vectors (\circled{7}) for validation.
Next, the \textbf{CGRA Mapping} phase maps the extracted DFG onto the CGRA fabric to maximize parallelism by exploiting intra- and inter-iteration parallelism with software pipelining (i.e., modulo scheduling)~\cite{software_pipelining}, and output compute and routing configurations.
Finally, in the \textbf{RTL Generation \& Verification} phase, the Hardware Generator (\circled{6}) uses the ADL model to produce Verilog RTL, which the Simulation/Emulation stage (\circled{8}) drives with the test vectors to verify functional correctness and collect area and power estimates.

\subsubsection{Architectural Specification \& Data-Flow Extraction}
\begin{figure}[h!]
    \centering
    \resizebox{\columnwidth}{!} {
    \includegraphics[width=\textwidth]{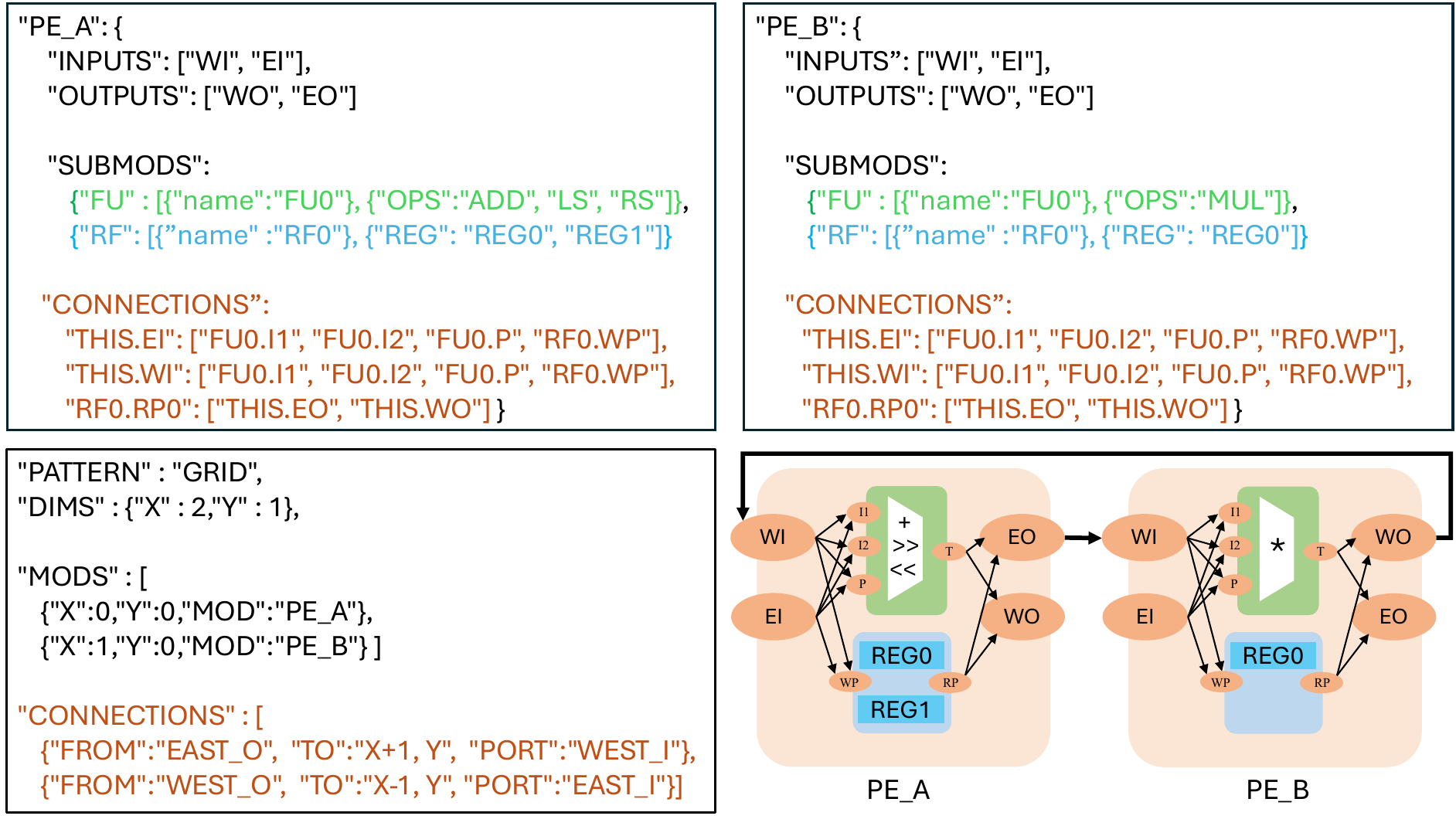}
    }
    \caption{Example of Morpher ADL for a heterogeneous CGRA with two processing elements. Internal connections of primitive modules (RF, FU) are omitted for simplicity.}
    \label{fig:adl_overview}
\end{figure}
\textbf{Architectural Description Language (ADL):}
Morpher’s ADL provides a concise yet expressive syntax for describing arbitrary CGRA architectures by defining three core abstractions: Modules (hardware blocks such as processing elements (PEs), register files, and memories), Ports (interfaces for connecting producers and consumers), and Connections  (interconnect wiring between Ports, including single‑cycle, multi‑hop links).  
Generally, all CGRAs can be described by hierarchically structuring the three Morpher ADL's primitive modules: Functional Units (FU), Register Files (RFs), and Memory Units (MU). Multiplexers are are automatically inferred based on the connections between ports. Fig.~\ref{fig:adl_overview} illustrates this by showing two heterogeneous PE instances, each composed of an FU supporting different operations, an RF, and dedicated input and output ports, linked together.

\textbf{DFG and Data Layout Generation:}
Morpher’s compiler frontend takes annotated C code as input 
and generates a Dataflow Graph (DFG) as output. The C code can come directly from the application or from DSLs like Triton~\cite{triton} or TVM~\cite{tvm} that can lower to C. Each node in the DFG represents a compute, memory, or predication operation, and includes all relevant metadata, ranging from the opcode and ASAP/ALAP hints for scheduling, to the number of parent nodes and whether the application exhibits any recursive behavior. Each node encodes information about its parent and child nodes as metadata. Depending on the memory access model supported by the target architecture, the DFG generation can be tailored to produce either on-array address computations or decoupled access-execute style addressing, as in the stream-dataflow model~\cite{softbrain} with explicitly orchestrated scratchpad banks~\cite{buffets_eddo}.

Once the DFG is constructed, Morpher also manages data layout into the CGRA’s memory banks. For example, in a multi-bank CGRA, it employs simple heuristics like round-robin allocation to minimize contention and memory bank conflicts. Finally, it emits a layout file that records base addresses for arrays and fixed locations for scalars, embedding this address metadata as constants into the corresponding DFG nodes. This data-rich DFG is then passed to the mapper, which parses it to schedule operations onto the actual CGRA fabric.

\subsubsection{CGRA Mapping}
The CGRA Mapper consumes the DFG and the architecture description to generate mapping configurations that minimize the initiation interval (II). II is defined as the number of cycles between the start of consecutive loop iterations and is constrained by resource availability and data dependencies. Starting from the theoretical Minimum II (MII), computed from resource availability and recurrence dependencies, the mapper attempts scheduling with II=MII and increments II until a feasible mapping is found. It first analyzes connectivity constraints between MUs and FUs, annotating each FU with the set of variables it can access. The DFG nodes are then ordered topologically, with recurrence‑cycle nodes prioritized by cycle length. Each node is placed onto a space–time instance of the corresponding FU in the Modulo Routing Resource Graph (MRRG), and routed from its parent nodes using Dijkstra’s shortest‑path algorithm. Ports may be temporarily oversubscribed to improve II convergence.

Once an initial mapping is obtained, the mapper resolves oversubscriptions through one of three strategies: an adaptive heuristic that increases the cost of overused ports (inspired by SPR), a simulated annealing (SA) approach that perturbs node placements along a cooling schedule, or a learning‑induced method (LISA) that leverages labels from a trained graph‑neural network. The process iterates until all resource conflicts are eliminated. Morpher’s modular design makes it straightforward to integrate new mapping algorithms, and future work will extend hierarchical and heterogeneous mapping techniques to improve scalability and support advanced CGRA topologies.

\subsubsection{Simulation, RTL Generation \& Verification}
\begin{figure}[h!]
    \centering
    \resizebox{0.9\columnwidth}{!} {
    \includegraphics[width=\textwidth]{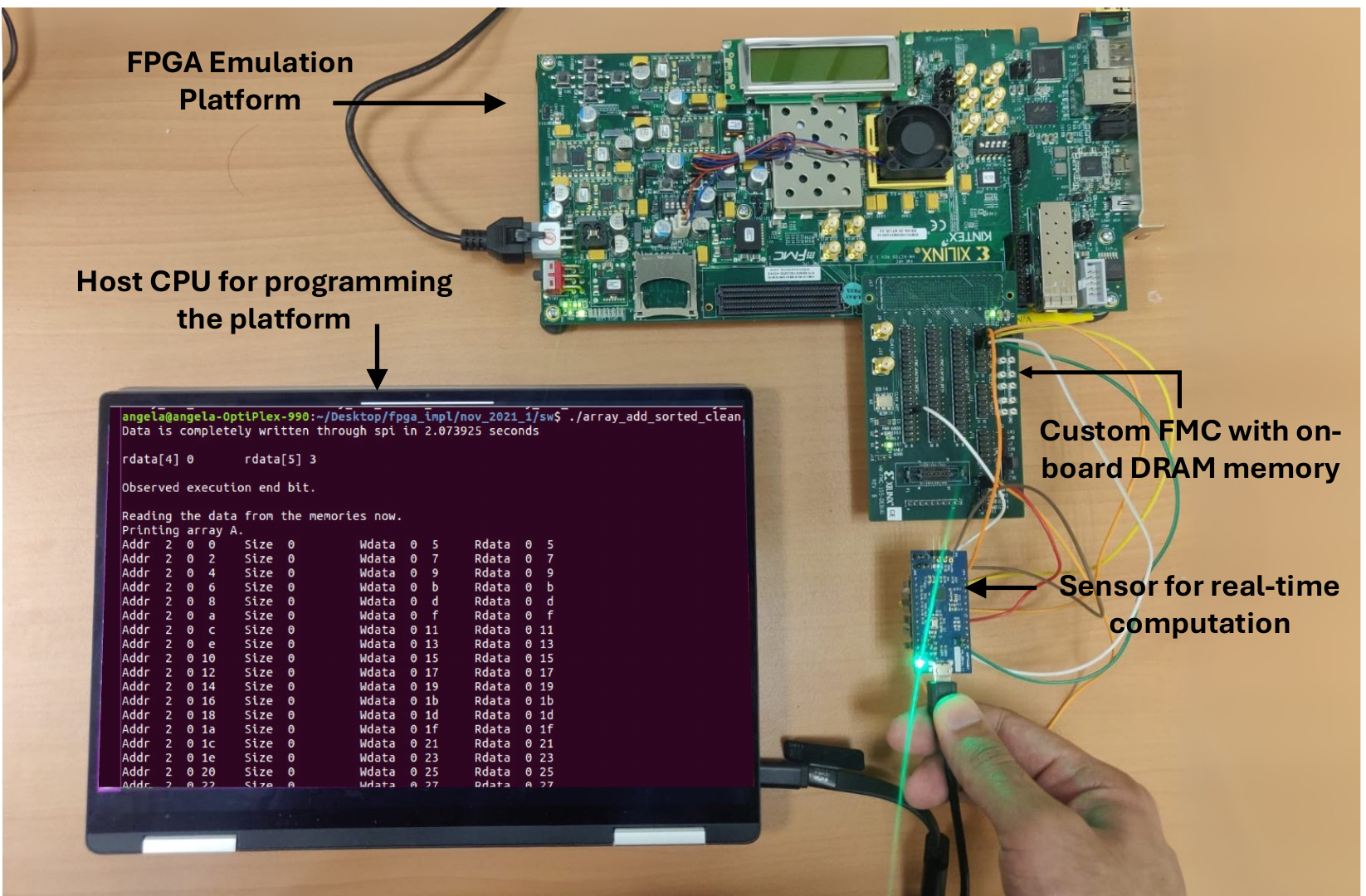}
    }
    \caption{FPGA emulation of PACE SoC 
    }
    \label{fig:emulation}
\end{figure}
Morpher’s ADL is also integrated with a simulation infrastructure, where the simulator parses the ADL and models the corresponding architecture. The simulator accepts the same bitstream and memory layout of the mapped kernel as a prototype CGRA as the input. This setup enables rapid evaluation of both performance and functional correctness of the mappings produced by the compiler.

Finally, Morpher supports RTL generation from the input JSON architecture file. For this, Morpher builds on Pillars~\cite{pillars}, utilizing its modular, Chisel-based infrastructure. The JSON file is parsed to instantiate the appropriate Chisel modules, which are then assembled into a complete hardware description and compiled into RTL. This RTL serves as a useful first-pass representation for early-stage area and power estimations, and enables design-space exploration within the Pareto-optimal envelope. However, additional effort is needed to refine this RTL into a concrete and verified implementation suitable for eventual tape-out. Morpher’s generated RTL currently supports out-of-the-box FPGA emulation for minor variants of the HyCUBE design. Fig.~\ref{fig:emulation} shows the Morpher-based emulation of a CGRA running a speech detection algorithm on an FPGA with sensors and controlled by a CPU. 
Nevertheless, automatic generation of valid RTL for arbitrary architectures remains dependent on human intervention, as it is an open challenge in design automation.

\subsection{HyCUBE: CGRA with single-cycle multi-hop interconnects}
\begin{figure}[h!]
    \centering
    \resizebox{0.9\columnwidth}{!} {
    \includegraphics[width=\textwidth]{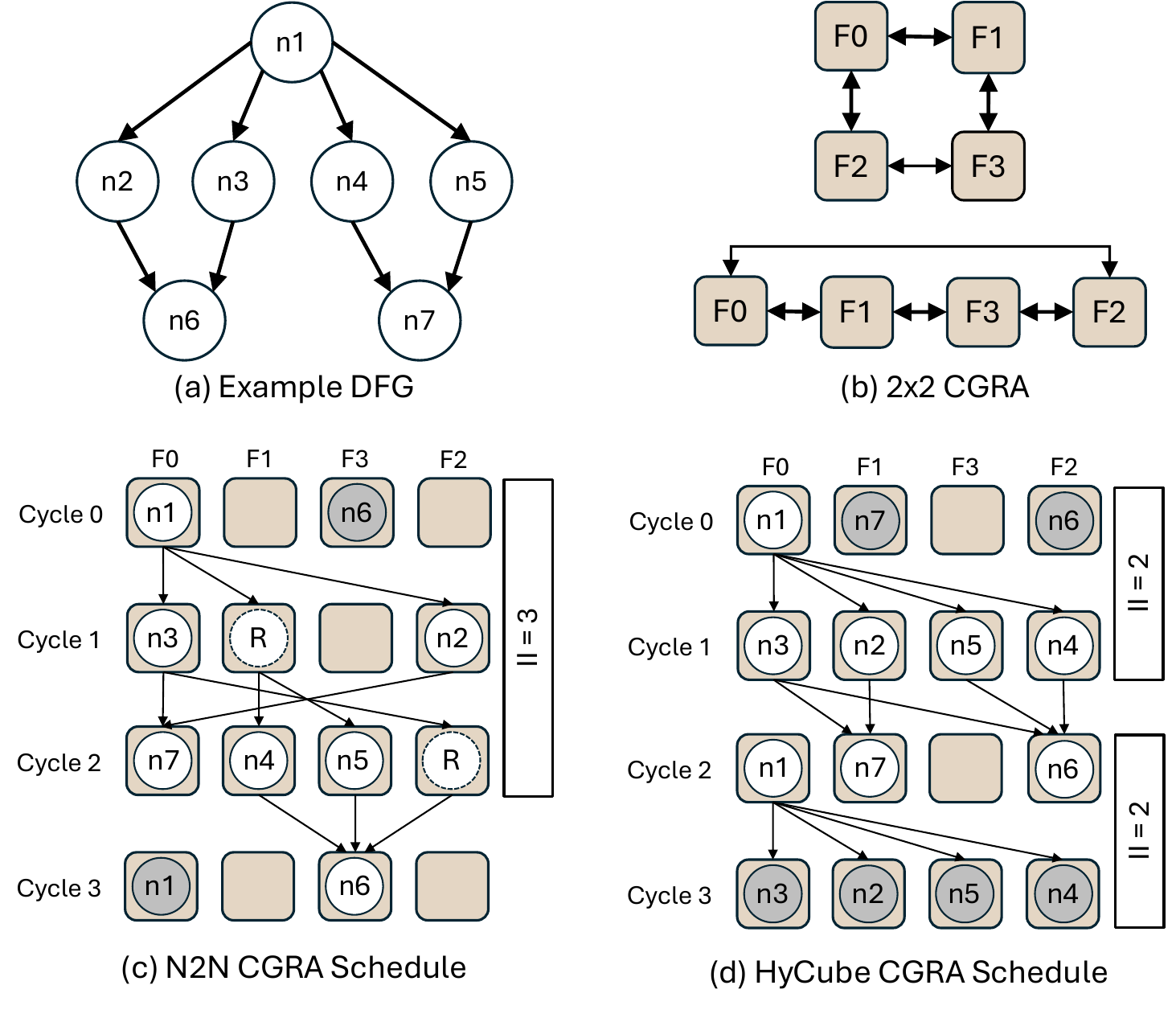}
    }
    \caption{Mapping of the DFG onto N2N and HyCUBE CGRAs; colored nodes in (c) and (d) indicate operations from the next loop iteration.\protect\footnotemark}
    \label{fig:hycube_schedule}
\end{figure}
\footnotetext{Adapted from HyCUBE~\cite{hycube}.}
Traditional CGRAs typically use neighbor-to-neighbor connectivity, where each processing element (PE) can only send data to its immediate north, south, east, or west neighbor. While this keeps the interconnect simple, it severely limits communication flexibility. If an operation needs to send data to a PE more than one hop away, the data must be routed through multiple intermediate PEs, one hop per cycle. 
These intermediate PEs are congested for routing, even though they're not performing any computation, effectively reducing the available compute parallelism. This type of routing increases the II. Fig.~\ref{fig:hycube_schedule}(c) illustrates a small loop kernel whose DFG is mapped onto a 2×2 N2N CGRA. Operation \textit{n1} is scheduled on tile \textit{F0} in cycle 0. Its dependents (\textit{n2}, \textit{n3}, \textit{n5}, \textit{n6}) are scheduled over the next few cycles. Due to limited connectivity, the compiler uses F1 and F2 to route the output of \textit{n1} to other PEs. This increases routing pressure, extends the schedule to three cycles per iteration (II = 3), and leaves fewer PEs for actual computation. 

In contrast, HyCUBE, shown in Fig.~\ref{fig:hycube_schedule}(d), introduces a compiler-scheduled, reconfigurable interconnect capable of single-cycle, multi-hop communication. Using clockless repeaters and statically configured crossbars, HyCUBE allows the compiler to set up paths that traverse multiple hops in a single cycle without occupying intermediate PEs. In the same DFG mapped to HyCUBE, \textit{n1} is still on F0 in cycle 0, but now all its dependent operations \textit{n2, n3, n5, and n6} can be scheduled in cycle 1. For example, the edge \textit{n1} to \textit{n5} is routed through the path PEs 0, 1 and 3 in a single cycle. Because the interconnect supports multicast, the same data can be sent to multiple destinations simultaneously without duplication. As a result, the II drops to 2 cycles, improving throughput and PE utilization. HyCUBE was fabricated in a commercial 40nm technology~\cite{hycube_tapeout}; our test chip delivers a peak energy efficiency of 26.4 MOPS/mW and consumes only 290 pJ per operation. The RTL of HyCUBE is open-sourced to enable further exploration and insight into the architecture.


HyCUBE solves a fundamental bottleneck in CGRA design. \textbf{It decouples routing from computation and gives the compiler fine-grained, cycle-level control over communication paths, all while preserving the energy and area efficiency that makes CGRAs attractive in the first place.}

\subsubsection{Microarchitectural Features}
HyCUBE features a 4×4 array of PEs, each equipped with an ALU, local configuration memory, and a crossbar switch. The leftmost column consists of memory-capable tiles that, in addition to computation, include load-store units (LSUs) connected to a shared 4-port data memory, while the remaining tiles are compute-only. The architecture’s key innovation lies in its compiler-scheduled interconnect: every PE’s crossbar output is driven by clockless repeaters that can be statically configured to bypass or latch data across directions (N, E, W, S). This allows data to travel across multiple PEs in a single cycle without using the PEs in between for routing. The interconnect supports multicast from a single source to multiple destinations within a cycle. This results in an extremely lightweight interconnect, made possible by relying entirely on compiler-determined routes and avoiding the complexity of dynamic routing or flow-control mechanisms. All connectivity is reconfigured cycle-by-cycle via instructions stored in each FU’s local configuration memory, making the interconnect highly efficient in both area and power.

HyCUBE eliminates the need for a centralized register file by using distributed registers placed at each directional input. Operands are sent straight to the ALU inputs, or temporarily stored at input registers if needed, which avoids extra move instructions and simplifies control. HyCUBE also supports predication for handling control divergence without requiring explicit predicate registers. Each ALU includes three input registers, two for operands and one for the predicate signal, and each operand includes a predicate flag. Execution occurs only when both predicate in the operands and the predicate input evaluate as valid, and downstream SELECT operations resolve control paths. To support cycle-accurate control, HyCUBE adopts a statically scheduled loop execution model. Thus, each PE's configuration memory is required to store one instruction per cycle over the II, with instruction contents specifying ALU operations, crossbar settings, register accesses, and constants necessary for correct execution.

\subsubsection{Compiler Support}
HyCUBE’s reconfigurable interconnect architecture necessitates a compiler that considers cycle-level communication paths in addition to traditional operation scheduling. In modulo scheduling for CGRAs, the compiler constructs a MRRG by unrolling the spatial architecture over a candidate II, replicating FUs, registers, and interconnects across time steps. The MII is computed as the maximum of the resource-constrained (ResMII) and recurrence-constrained (RecMII) bounds~\cite{recmii}. In traditional CGRAs, data dependencies between distant FUs are mapped as multi-cycle paths through intermediate FU nodes, which are temporarily occupied for routing. In contrast, HyCUBE introduces cycle-level reconfigurability in the interconnect, requiring the MRRG to explicitly model links between FUs as schedulable resource nodes. These links are used for single-cycle, multi-hop paths that connect FUs mapping source and sink nodes directly, without involving intermediate FUs.

HyCUBE leverages key components of the Morpher's infrastructure for its compiler-to-hardware integration. The architecture is described using Morpher’s ADL primitives. 
HyCUBE also provides a custom ISA specification, which Morpher uses to automatically generate cycle-accurate configurations and bitstreams for each PE, along with test vectors for functional simulation. Morpher further outputs initial RTL based on parameterized modules, which is used for early evaluation of area and power metrics through commercial synthesis tools. While final deliverable RTL and full backend implementation for tapeout required additional engineering effort, Morpher’s modular infrastructure significantly accelerated initial development and system-level validation.

\subsection{PACE: RISC-V SoC with integrated CGRA}
\begin{figure}[h!]
    \centering
    \resizebox{\columnwidth}{!} {
    \includegraphics[width=\textwidth]{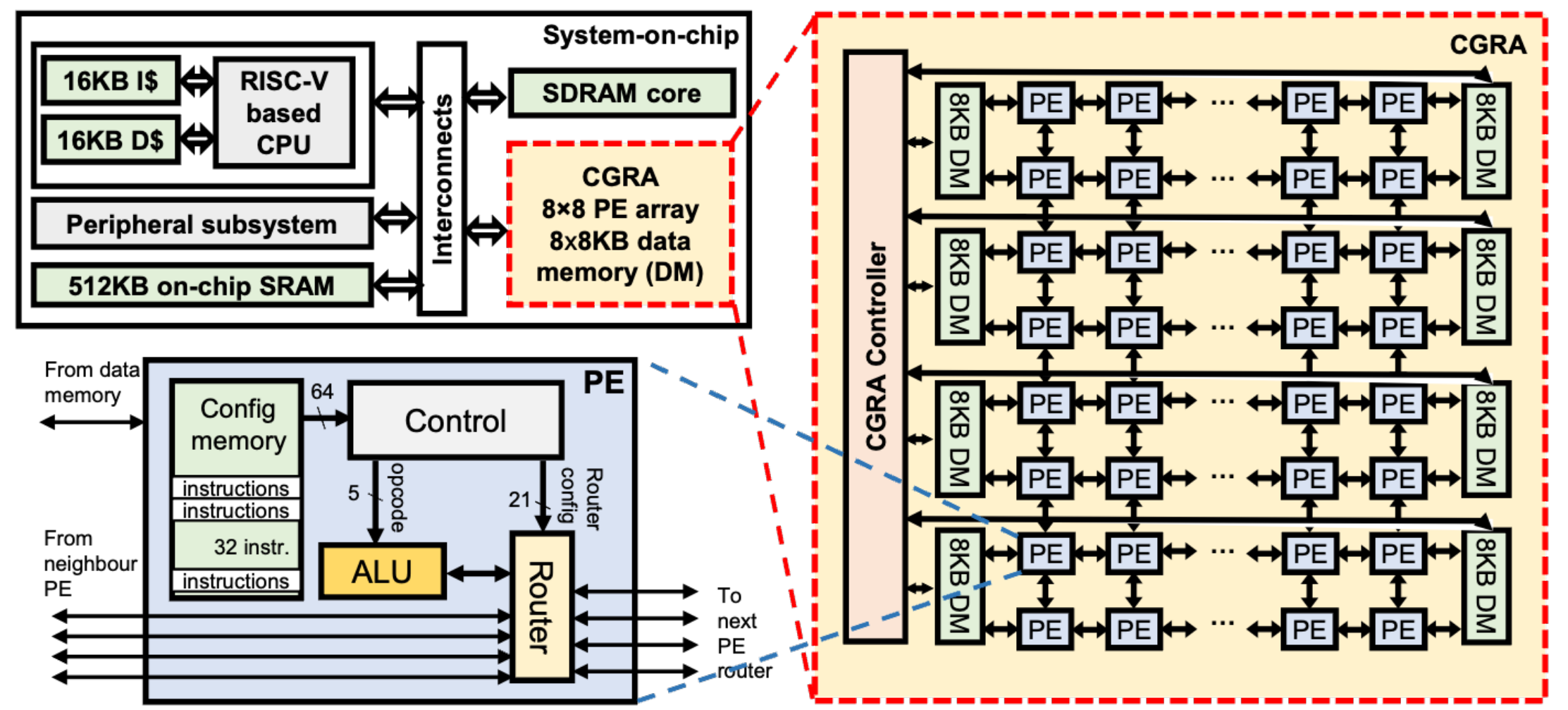}
    }
    \caption{PACE: CGRA integrated in a RISC-V system-on-chip (SoC)\protect\footnotemark}
    \label{fig:pace_arch}
\end{figure}
\footnotetext{Adapted from PACE~\cite{pace, pace_isocc}.}

\begin{figure*}[h!]
    \centering
    \resizebox{\textwidth}{!} {
    \includegraphics[width=\textwidth]{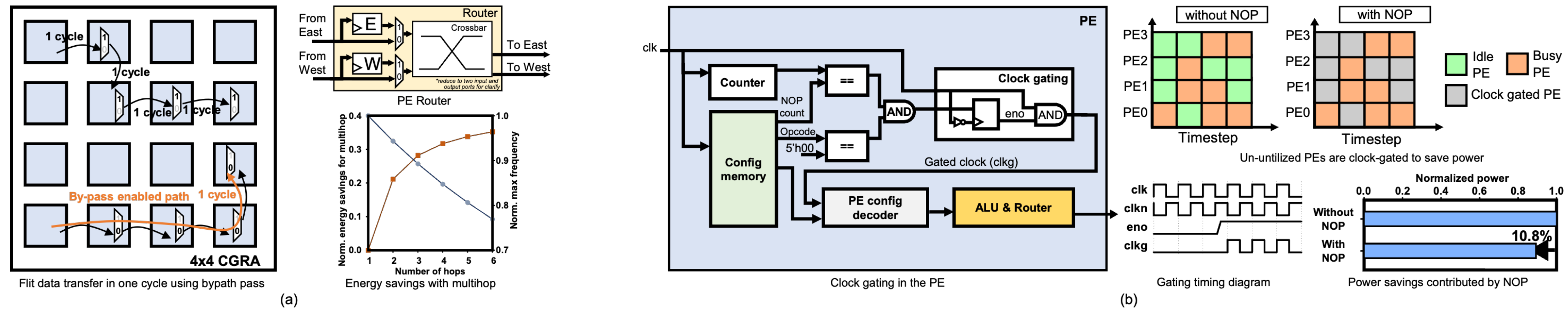}
    }
    \caption{(a) PE router with bypath pass for single-cycle multi-hop data. (b) Dynamic clock gating for power savings.\protect\footnotemark[\value{footnote}]}
    \label{fig:pace_router}
\end{figure*}

\begin{figure}[h!]
    \centering
    \resizebox{\columnwidth}{!} {
    \includegraphics[width=\textwidth]{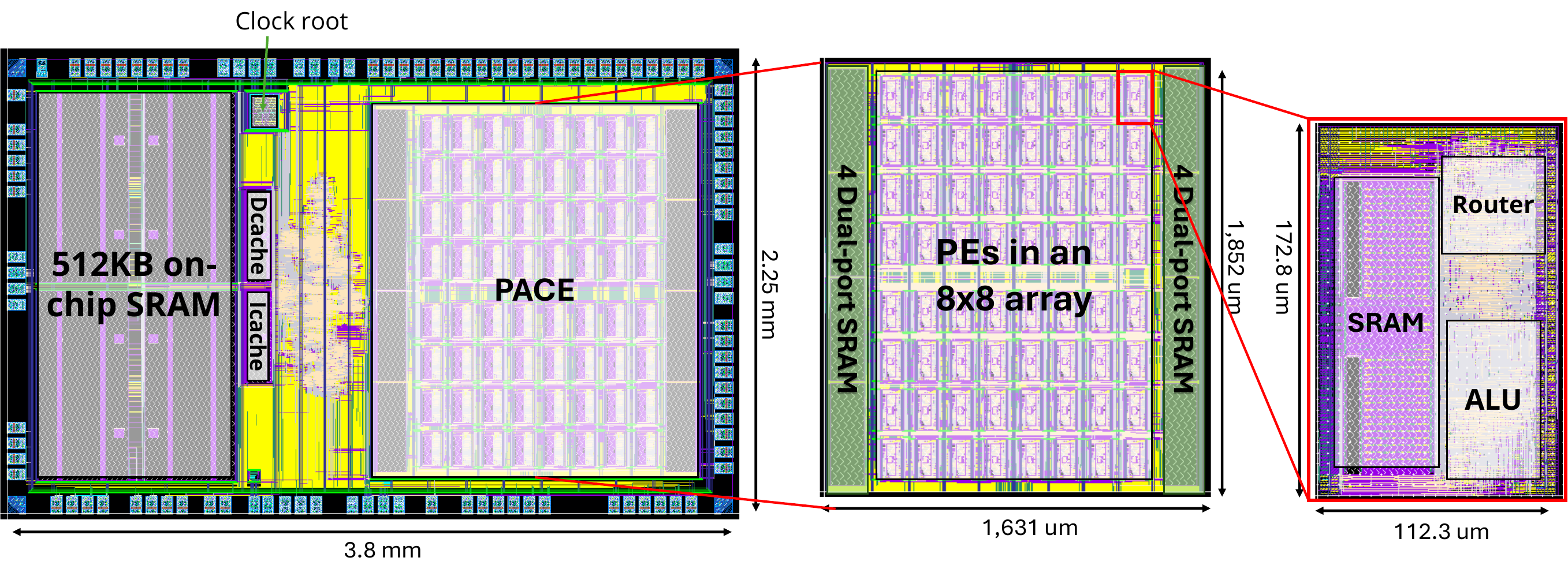}
    }
    \caption{Modular PACE chip layout, with compact PE layout.\protect\footnotemark[\value{footnote}]}
    \label{fig:pace_tapeout}
\end{figure}

PACE~\cite{pace, pace_isocc} is a fabricated chip implemented in a commercial 40nm technology node, achieving a peak efficiency of 360 GOPS/W at 0.6V. It extends the HyCUBE architecture by scaling from a single 4×4 CGRA cluster to a 64-PE (8×8) CGRA integrated within a RISC-V-based SoC. The PACE SoC features a 32-bit RISC-V core that manages execution by coordinating data transfers to the CGRA’s data and configuration memories, and updates control registers through interrupt-based signaling. Each PE features a 16-bit ALU, 0.25KB of compiler-managed configuration memory, and a statically scheduled, multi-hop crossbar interconnect. Compared to HyCUBE’s 32-bit datapaths, all PEs in PACE now use 16-bit datapaths, better aligned with modern AI workload requirements. The CGRA is partitioned into four interconnected clusters, enabling efficient and seamless data communication across cluster boundaries.

To support real-world embedded workloads, the PACE SoC integrates a broad set of external peripherals and memory interfaces. As shown in Fig.~\ref{fig:pace_arch}, the SoC includes an on-chip SRAM and a memory controller that interfaces with an external 32MB SDRAM chip for handling larger data storage requirements. For peripheral interfacing, the SoC includes standard communication protocols such as UART, SPI, and I2C which connect to off-chip components including a bluetooth module, SD card controller, and I2C ROM. Additionally, dedicated ports are provided for general-purpose I/O (GPIO), analog-to-digital conversion (ADC), and cryptographic acceleration via AES and RNG modules. These components are accessed through a shared AXI4 slave multiplexer, providing a uniform MMIO interface to the system. This rich peripheral set enables PACE to support diverse I/O and memory-intensive tasks, making it suitable for real-world edge computing workloads.

PACE further improves its energy efficiency compared to HyCUBE by employing both static and dynamic clock gating. Static clock gating is applied at compile time by identifying periods where specific PEs are idle and disabling their clocks. Complementing this, dynamic clock gating contributes to an additional 10\% power reduction through the insertion of idle-state instructions and associated gating logic, as shown in Fig.~\ref{fig:pace_router}. The compiler inserts NOP instructions to specify intervals during which certain PEs remain idle. These instructions encode the start and end times, allowing local counters within each PE to track the idle period precisely. When the counter is active, the clock to most of the PE’s internal logic is gated, excluding critical components like the routing logic that must remain functional. After the counter expires, the PE resumes execution as scheduled. This compiler-coordinated approach allows fine-grained, cycle-accurate clock control that exploits the dataflow execution model, maximizing energy savings.

\subsection{Morpher has enabled broader research in CGRAs}
The CGRA design space remains rich with open problems and optimization opportunities across the stack ranging from improved DFG generation tailored to memory access patterns, to enhanced backend compilation involving mapping, placement, and routing, and even the design of novel architectures better suited to specific classes of kernels. Morpher has, in part, enabled several research efforts across this spectrum.

REVAMP~\cite{revamp} is a design-space exploration framework that leverages Morpher’s ADL to instantiate heterogeneous CGRA configurations. It employs Morpher’s DFG generator to extract dataflow from annotated C code, and extends Morpher’s compiler backend with its own customized toolchain. LISA~\cite{LISA} builds on Morpher’s compilation flow by replacing its simulated annealing-based mapper with a GNN-based labeling and mapping strategy, generalizable to a range of spatial accelerators and CGRAs. Nexus Machine~\cite{nexus} and Canon~\cite{canon} both utilize Morpher’s DFG generation infrastructure to compile arbitrary application kernels for their respective fabrics. Morpher and HyCUBE have also provided impetus to other CGRA architectures~\cite{snafu, riptide, cascade_jackson}, enabling exploration of ideas such as multi-hop routing to improve compilation efficiency and flexibility on their fabrics.
CTScan~\cite{ctscan} is another interesting work that leverages CGRAs to emulate power side-channels of edge CPUs. It leverages multi-hop capabilities of the PACE CGRA chip to emulate data forwarding across CPU pipeline stages. Prior works in CGRAs~\cite{snafu, softbrain, cascade_jackson} have been modeled in Morpher for better insight into various architectural features. Broadly, Morpher’s DFG generation, compilation flow, and ADL form a robust research infrastructure, enabling researchers to benchmark, ablate, and contrast their contributions or innovations against prior work.
\section{Experimental Study}

This study demonstrates how Morpher can be used to not only accelerate kernels on CGRA-based systems but also is supporting architecture design space exploration. It also highlights the importance of supporting control divergence and recurrence edges. 

\subsection{Mapping Quality of Morpher}
\textbf{Benchmark Kernels:}
We use kernels from a range of popular benchmark suites, including MachSuite, Polybench, Wavelib, and BEEBS, spanning multiple domains and comprising commonly used edge kernels for image processing, filtering, machine learning, and basic linear algebra. Benchmarks with a small number of graph nodes can be mapped onto the spatial CGRA without partitioning; for such cases, we additionally evaluate unrolled variants (\textit{\_u}) to provide better insights.


\textbf{Baseline Architectures:} We evaluate spatial and spatio-temporal CGRA variants from the earlier taxonomy in Fig.~\ref{fig:ii_result}. The spatial and spatio-temporal architectures are modeled after Snafu~\cite{snafu} and HyCUBE~\cite{hycube}, respectively. Across all benchmarks, the spatial architecture exhibits an equal or higher II than the spatio-temporal counterpart, trading some performance for lower power by eliminating configuration memory. Table~\ref{tab:hop_result} quantifies the impact of introducing multi-hop interconnects into CGRA compilation. With two hops, the CGRA already shows performance gains across benchmarks; with four hops, the improvement frequently exceeds 50\%, owing to the additional routing freedom and flexibility.

\begin{figure*}[h!]
    \centering
    \resizebox{0.9\textwidth}{!} {
    \includegraphics[width=\textwidth]{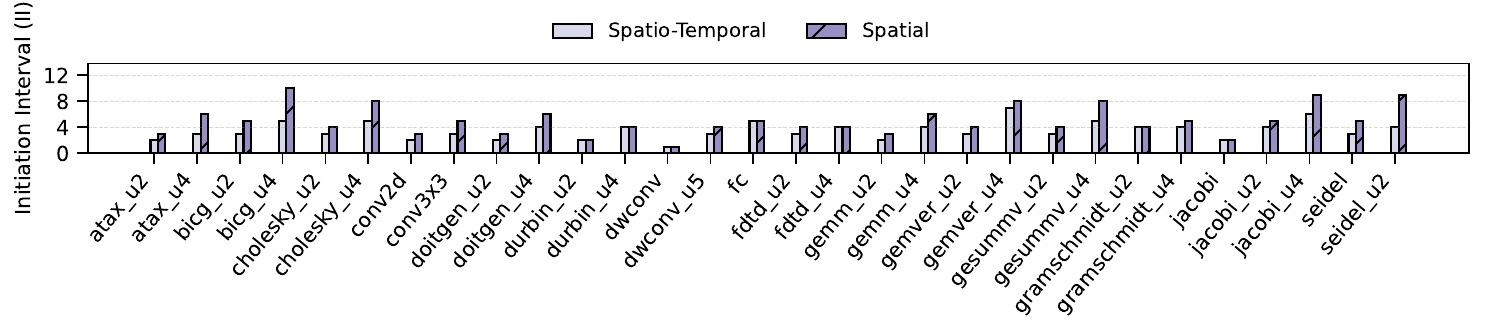}
    }
    \caption{Summary of benchmark mapping results for spatial and spatio-temporal CGRA architectures.}
    \label{fig:ii_result}
\end{figure*}

\begin{table}
    \centering
    \begin{tabular}{|c|c|c|c|c|}\hline
         \textbf{App. Kernels}&  \textbf{1-hop}&  \textbf{2-hops}&  \textbf{3-hops}& \textbf{4-hops}\\\hline
         \textbf{fft}        &  11&  5&  5& 5\\\hline
         \textbf{adpcm}&  17&  9&  9& 8\\\hline
         \textbf{aes}        &  24&  15&  13& 13\\\hline
         \textbf{disparity}&  26& 12&  10& 11\\\hline
         \textbf{dct}&  23&  14&  13& 13\\\hline
 \textbf{nw}& 19& 15& 15&15\\\hline
 \textbf{GeMM}       & 14& 9& 8&7\\ \hline
    \end{tabular}
    \caption{Impact of multi-hop interconnects on CGRA performance.}
    \label{tab:hop_result}
\end{table}

\subsection{PACE SoC evaluations}
\begin{figure}[h!]
\resizebox{\columnwidth}{!} {
    \centering
    \begin{subfigure}[b]{0.32\textwidth}
        \centering
        \includegraphics[width=\textwidth]{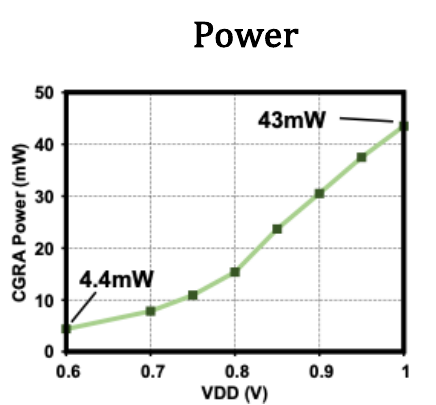}
        \caption{}
        \label{fig:a}
    \end{subfigure}
    \hfill
    \begin{subfigure}[b]{0.32\textwidth}
        \centering
        \includegraphics[width=\textwidth]{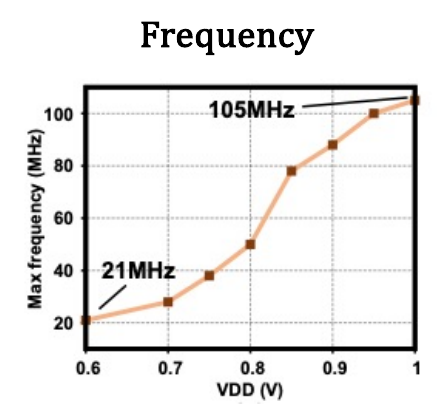}
        \caption{}
        \label{fig:b}
    \end{subfigure}
    \hfill
    \begin{subfigure}[b]{0.32\textwidth}
        \centering
        \includegraphics[width=\textwidth]{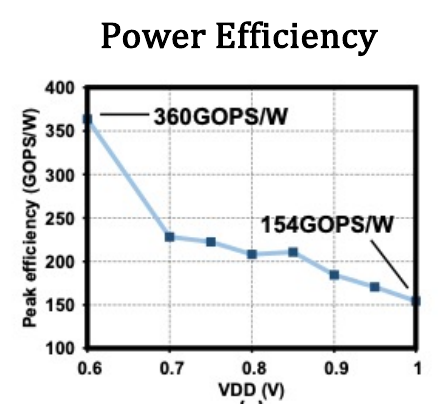}
        \caption{}
        \label{fig:c}
    \end{subfigure}
}
    \caption{(a) CGRA power vs. VDD, (b) max frequency vs. VDD,  (c) power efficiency vs. VDD.}
    \label{fig:pace_frequency_power}
\end{figure}

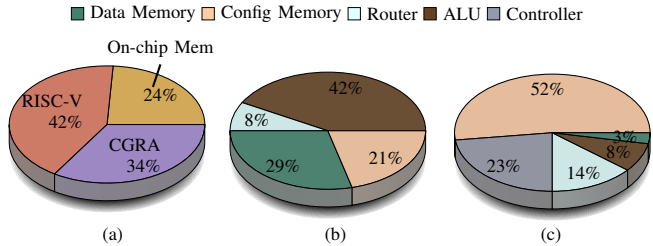
\begin{figure}[t] 
	\scriptsize
	\centering
    \captionsetup[subfloat]{labelfont=scriptsize,textfont=scriptsize}
    \subfloat{
	\begin{tikzpicture}[
  recofig/.style={shape=rectangle, draw=black, fill=deeppeach, line width=0.1},
  other/.style={shape=rectangle, draw=black, fill=manatee,   line width=0.1},
  compute/.style={shape=rectangle, draw=black, fill=darkbrown,line width=0.1},
  router/.style={shape=rectangle, draw=black, fill=lightcyan, line width=0.1},
  dataMem/.style={shape=rectangle, draw=black, fill=viridian,  line width=0.1}]
  \matrix [node font=\scriptsize] at (-2,0) {
    \node [dataMem,label=right:Data Memory] {}; &
    \node [recofig,label=right:Config Memory] {}; &
    \node [router,label=right:Router] {}; &
    \node [compute,label=right:ALU] {}; &
    \node [other,label=right:Controller] {}; \\
  };
\end{tikzpicture} 
    }
    \\
    \setcounter{subfigure}{0}
    \subfloat[]{
\begin{tikzpicture}
\definecolor{onchipmem}{HTML}{E6B450} 
\definecolor{riscv}{HTML}{E07A5F}     
\definecolor{cgra}{HTML}{A48AD4}     

\piechartthreed[scale=0.26, background color=orange!50, mix color=gray]
  {86.4/onchipmem, 151.2/riscv, 122.4/cgra}

\node[align=right] at ($(pc 3) + (-0.1, 0.1)$) {CGRA\\34\%};

\node[align=right] at ($(pc 2) + (0.2,0)$) {RISC-V\\42\%};

\coordinate (eL3) at ($(pc 1) + (0, 0.40)$); 
\draw[-, thick] ($(pc 1)+(-0.1,0.1)$) -- (eL3)
    node[above, align=center, rotate=0] {On-chip Mem}; 
\node at ($(pc 1)$) {24\%}; 

\end{tikzpicture}
    \label{fig:a}
    }
    \subfloat[]{
    \hspace{-0.3cm}
\begin{tikzpicture}
\piechartthreed[scale=0.26, background color=orange!50, mix color=gray]
  {151.2/darkbrown, 28.8/lightcyan, 104.4/viridian, 75.6/deeppeach}

\draw[black] (pc 1) node{42\%};

\draw[black] (pc 2) node{8\%};

\draw[black] (pc 3) node{29\%};

\draw[black] (pc 4) node{21\%};
\end{tikzpicture}
    \label{fig:b}
    }    
    \subfloat[]{
\begin{tikzpicture}
\piechartthreed[scale=0.26, background color=orange!50, mix color=gray]
  {187.2/deeppeach, 82.8/manatee, 50.4/lightcyan, 28.8/darkbrown, 10.8/viridian}

\draw[black] (pc 1) node{52\%};

\draw[black] (pc 2) node{23\%};

\draw[black] (pc 3) node{14\%};

\draw[black] (pc 4) node{8\%};

\draw[black] (pc 5) node{3\%};
\end{tikzpicture}
    \label{fig:c}
    }   
	\caption{Area breakdown of (a) the PACE SoC, (b) the CGRA, and (c) power breakdown of CGRA.\protect\footnotemark[\value{footnote}]} 
	\label{fig:power_area}
\end{figure}

We measured the PACE silicon across 0.6-1.0 V. CGRA power scales from 4.4 mW at 0.6 V to 43 mW at 1.0 V as shown in Fig.~\ref{fig:pace_frequency_power}(a), while the maximum clock increases from 21 MHz to 105 MHz (Fig.~\ref{fig:pace_frequency_power}(b). Energy efficiency peaks at 360 GOPS/W at 0.6 V/21 MHz and decreases toward $\sim$154 GOPS/W near 0.95-1.0 V as dynamic power grows faster than throughput (Fig.~\ref{fig:pace_frequency_power}(c). These curves illustrate the expected energy-performance trade-off and motivate operating near 0.6-0.7 V for energy-limited edge deployments, or near 0.9-1.0 V when throughput dominates.

The fabricated SoC (40 nm) occupies 7.6 mm$^2$, as shown in Fig.~\ref{fig:pace_tapeout}. As shown in Fig.~\ref{fig:power_area}(a), system-level area splits into RISC-V controller 42\%, on-chip SRAM 24\%, and CGRA 34\%. Within the CGRA, area is dominated by PE logic (ALU + control) at 42\%, followed by data memory at 29\%, PE configuration memory (CM) at 21\%, and routing at 8\% (Fig.~\ref{fig:power_area}(b). Power attribution shows a different balance: CM accounts for 52\% of CGRA power, with PE controller 23\%, router 14\%, ALU 8\%, and data memory 3\% (see Fig.~\ref{fig:power_area}(c). Configuration memory, though modest in area, consumes the most power because the CM is read every cycle to configure ALU, router, and register controls for all PEs.

\begin{table}[h]
\centering
\resizebox{\columnwidth}{!} {
\begin{tabular}{|l|c|c|c|c|c|}
\hline
 & \textbf{Amber \cite{amber}} & \textbf{SSCL \cite{sscl}} & \textbf{ISSCC \cite{isscc}} & \textbf{JSSC \cite{jssc}} & \textbf{PACE (Our work)} \\ \hline
\textbf{Year} & 2022 & 2020 & 2019 & 2020 & 2023 \\ \hline
\textbf{Tech (nm)} & 16 & 28 & 22 & 28 & 40 \\ \hline
\textbf{Area (mm$^2$)} & 20.1 & 3.9 & 4.9 & 4.80 & 3.02 \\ \hline
\textbf{\#PEs} & 384 & 120 & 15 & 64 & 64 \\ \hline
\textbf{Voltage (V)} & 1.29 & 0.6 & 0.48 & 0.9 & 0.6 \\ \hline
\textbf{Freq (MHz)} & NA & 89 & 46 & 800 & 21 \\ \hline
\textbf{Power (mW)} & NA & 45.9 & NA & 537 & 4.4 \\ \hline
\textbf{Efficiency (GOPS/W)} & 538 & 307 & 978 & 196 & 360 \\ \hline
\textbf{Memory} & 4500KB & 234KB & 690KB & 320KB & 80KB \\ \hline
\textbf{Norm. area (mm$^2$)$^{1}$} & 50 & 5.5 & 3.2 & 6.86 & 3.02 \\ \hline
\textbf{Norm. efficiency (GOPS/W)$^{2}$} & 86 & 150 & 296 & 96 & 360 \\ \hline
\end{tabular}
}
\vspace{1ex}
\raggedright
\footnotesize
$^{1}$ Norm. area = Area $\cdot \frac{\text{node}}{40\text{nm}}$;  
$^{2}$ Norm. efficiency = Efficiency $\cdot \left( \frac{\text{node}}{40\text{nm}} \right)^2$.
\caption{Comparison of this work with prior designs\protect\footnotemark[\value{footnote}].}
\label{tab:comparison}
\end{table}

Table~\ref{tab:comparison} compares our CGRA with prior published designs, with efficiency and area normalized to 40 nm to account for process differences. To isolate architectural effects, we disabled workload-dependent power-saving features (e.g., idle-state clock gating) during measurements. Under these conditions, the CGRA achieves 360 GOPS/W at 0.6 V, exceeding prior work by $1.2\times$-$4.6\times$ on the normalized metric, and occupies 3.02 mm$^2$ (normalized) for the 64-PE array. Beyond peak numbers, the architecture and compiler support a broad mix of kernels and can execute multiple distinct kernels concurrently, indicating greater versatility.

\section{Toward A Unified Abstraction Layer \\ for Spatial Accelerators}

Morpher seeks to unify a landscape of highly specialized yet fragmented CGRA architectures, but a considerable amount of work remains to fully realize this vision. Its ADL, while detailed, must be extended to capture a wider spectrum of complex and programmable architectures. It remains constrained by a set of supported network topologies, PE types, and memory banking schemes. For instance, the current PE model is limited to a composition of an ALU, a router, and a register file, whereas more sophisticated architectures may incorporate PEs with various IP blocks, coarser-grained compute cores, caches, or more intricate interconnects. Moreover, the ADL’s verbosity imposes a barrier to intuition and creates friction for non-experts attempting to specify architectures.

Morpher framework also lacks programming abstractions that would allow both expert and non-expert users to manually explore architectural or mapping optimizations at a granularity aligned with their expertise. Moreover, its supported architectures primarily target kernels with regular computation patterns and employ a high control-to-compute ratio to preserve generality, resulting in inefficiency relative to domain-specific accelerators. 


These challenges motivates a vision for a unified abstraction layer that decouples hardware specialization from software development. In essence, we seek a generalized and common intermediate representation (IR) of spatial computation and communication that remains consistent regardless of the underlying accelerator, allowing software to target a virtual spatial architecture rather than a particular accelerator implementation.

Applications will be first lowered from high‑level languages or Domain-Specific Languages (DSL) into this virtual spatial IR. Subsequently the IR is mapped to a particular accelerator through target‑specific back ends that perform mapping, scheduling, and routing to generate device configurations/bitstreams. A common IR allows front ends and back ends to evolve independently. The back ends would be parameterized by a declarative architecture specification as an extension of Morpher’s ADL of modules, ports, and connections so that the same IR can be retargeted without per‑device rewrites.

An architecture independent intermediate representation (IR) for spatial computing would decouple software from device details and enable portability: developers can write kernels once (for example in Triton~\cite{triton}) and run them on many CGRAs and related accelerators that implement the common IR. Hardware designers can still innovate in PE organization, interconnects, and memory systems without rebuilding the software stack. This separation acts as a stable interface between software and hardware, built on a small set of composable spatial primitives with precise semantics, which extends the lifetime of compilers and tools~\cite{dsagen}. A shared IR would also focus compiler work on the common layer, instead of writing new mappers and schedulers for every architecture, as shown by MLIR dialects that decouple CGRA compilation from specific back ends~\cite{mlir-compile-cgra}. With this foundation, optimizations for dataflow orchestration, pipelining, and parallel scheduling can be written once and reused across devices.

More importantly, making the abstraction open and modular would establish a scalable foundation for spatial computing. An open standard invites community collaboration: academic and industry contributors could align on interface definitions, contribute to common libraries, and collectively drive improvements. Over time, such a foundation can evolve to incorporate new hardware capabilities (e.g., novel functional units or memory models) while preserving backward compatibility, much as extensible ISAs like RISC-V have done for general-purpose processors. We envision that embracing a unified, open abstraction layer will transform the spatial accelerator landscape into a more portable, innovative, and sustainable ecosystem, combining the efficiency of specialization with the flexibility of a shared, interoperable infrastructure.


\section{Acknowledgments}
This research is partially supported by the National Research Foundation, Singapore under its Competitive Research Programme Award NRF-CRP23-2019-0003.

\bibliographystyle{IEEEtrans}
\clearpage
\bibliography{refs.bib}

\end{document}